\documentclass[final,5p,times,twocolumn,nopreprintline]{elsarticle}
\usepackage{amsmath,slashed,booktabs}
\usepackage{graphicx,graphics}
\usepackage{dcolumn}
\usepackage[hyperfootnotes=false]{hyperref}
\usepackage{xspace}
\usepackage{color}
\usepackage{balance}
\usepackage{multirow}
\usepackage{multicol}

\usepackage{fancyhdr}
\fancypagestyle{firstpage}{%
	
	\lhead{}
	\rhead{\small INT-PUB-25-014}
}
\newcommand{\GeV}{\,\text{GeV}}
\newcommand{\MeV}{\,\text{MeV}}
\newcommand{\meV}{\,\text{meV}}

\newcommand{\eV}{\,\text{eV}}

\newcommand{\Order}{\mathcal{O}}
\newcommand{\beq}{\begin{equation}}
\newcommand{\eeq}{\end{equation}}

\allowdisplaybreaks[4]

\newcommand{\onbb}{0\nu\beta\beta}
\newcommand{\mbb}{|m_{\beta\beta}^{\rm eff}|}
\newcommand{\mb}{m_\beta}
\newcommand{\msum}{\sum_i m_{\nu_i}}
\newcommand{\slashpartial}{\protect{\slash\hspace{-0.5em}\partial}}
\newcommand{\diag}{\operatorname{diag}}
\renewcommand{\i}{\mathrm{i}}
\newcommand{\mvmin}{m_{\nu_{\rm min}}}
\newcommand{\Thalf}{T_{1/2}^{0\nu}}
\newcommand{\spacevec}[1]{\mathbf{#1}}

\newcommand{\typeIrefs}{\cite{Deppisch:2015qwa,Agostini:2017jim,Abdullahi:2022jlv,Antel:2023hkf}}

\begin{document}

\renewcommand{\theequation}{\arabic{equation}}

\begin{frontmatter}
 
\title{
Neutrino-less double beta decay in the $\nu$ Standard Model 
}

\author[Seattle]{Vincenzo Cirigliano}
\author[Seattle]{Wouter Dekens}
\author[Seattle,Evanston]{Sebastián Urrutia Quiroga}

\address[Seattle]{Institute for Nuclear Theory, University of Washington, Seattle, WA 91195-1550, USA}
\address[Evanston]{Northwestern University, Department of Physics \& Astronomy, 2145 Sheridan Road, Evanston, IL 60208, USA}

\begin{abstract}
We perform a comprehensive study of the $3+3$ Type-I seesaw model for a broad range of 
right-handed mass scales (from keV to 10 TeV).   We take into account and, in some cases, update the constraints from a large number of high- and low-energy experiments  
and study the implications on neutrino-less double beta ($0\nu\beta\beta$) decay experiments.  
We illustrate our findings through profile likelihood plots for the half-life $T_{1/2}^{0\nu}$ and two-dimensional plots correlating $T_{1/2}^{0\nu}$ to neutrino masses.   
We find that in this simple class of models for Majorana neutrino masses, current and next-generation $0\nu\beta\beta$ decay experiments have a broad discovery potential in both the normal and inverted orderings of the spectrum of light active neutrinos.        
\end{abstract}

\end{frontmatter}

\thispagestyle{firstpage}

\section{Introduction}
\label{sec:intro}

Understanding the origin and nature of neutrino mass is one of the forefront questions driving particle and nuclear physics.  
The inclusion of $N$ fermionic  Majorana fields that are singlets under the Standard Model (SM) gauge group, often dubbed as `right-handed neutrinos', `sterile neutrinos',  or `heavy neutral leptons', provides an attractive minimal model for the generation of neutrino mass (see \cite{Abdullahi:2022jlv} and references therein for a recent review).  
This simple scenario, with $N=3$, has the potential to 
explain other puzzles, such as the nature of dark matter and the origin of the 
cosmic baryon asymmetry~\cite{Asaka:2005pn,Canetti:2012kh}.
Moreover, one needs at least $N=3$ if the lightest neutrino mass is non-vanishing.

The mass scales associated with the eigenvalues of the right-handed neutrino mass matrix ($M_R$) are free parameters of the model. 
Here we take $N=3$ and assume that a hierarchy exists between the eigenvalues of $M_R$ and the masses of the light active neutrinos  
(without any additional restriction on $M_R$), \emph{i.e.,} we focus on the $3+3$ Type-I seesaw model. 
This setup provides perhaps the simplest scenario for the generation of neutrino masses, 
which 
are generically of Majorana nature. 

The Type-I seesaw model has been studied extensively in the literature. 
In the high-scale version of the model (with $M_R$ close to the GUT scale), the only low-energy remnant is a Majorana mass matrix for the light active neutrinos. 
If the right-handed mass is at the TeV scale or lower, a rich phenomenology opens up, and this has motivated a vibrant program of experimental searches for 
right handed neutrinos~\cite{Abdullahi:2022jlv}. 
In particular, given the Majorana nature of the resulting six massive neutrinos, 
in this framework, lepton number ($L$) is not conserved. In fact, this model induces a non-zero rate for neutrinoless double beta 
($0\nu \beta \beta$)  decay and predicts 
distinctive $\Delta L=2$ signatures at colliders.  
Yet, to our knowledge, a comprehensive analysis of the 
discovery potential and constraining power of current and next-generation 
searches for neutrinoless double beta ($0\nu \beta \beta$) decay is missing. 
Motivated by this,  in this manuscript  we revisit 
the experimental constraints on the $3+3$ Type-I seesaw model for 
a broad range of experimentally accessible right-handed mass scale (from keV to $10$ TeV)
{\it and} study the implications of all data on $\onbb$ decay.

The manuscript is organized as follows: in Section~\ref{sec:model} we describe the setup of the model, in Section~\ref{sec:scan} we summarize the experimental constraints, highlighting updates compared to the 
previous literature, and describe the parameter scan. 
In Section~\ref{sec:onbb} we summarize the   $0\nu \beta \beta$ decay rate 
formula appropriate for the $3+3$ scenario, before presenting our results in Section~\ref{sec:results} 
and conclusions in Section~\ref{sec:conclusions}.

\section{Model description}
\label{sec:model}
We focus on the minimal extension of the SM that involves, in addition to the SM fields, a set of three $\nu_{Ri}$ spin-1/2 singlets, the so-called sterile neutrinos. This scenario is often called $\nu$SM \cite{Asaka:2005pn} and can explain nonzero neutrino masses by the type-I seesaw mechanism \cite{Minkowski:1977sc, Gell-Mann:1979vob, Yanagida:1979as, Glashow:1979nm, Mohapatra:1979ia}. Including operators up to dimension-4, this extends the SM Lagrangian by a kinetic term, a Majorana mass term, and Yukawa interactions of the $\nu_{Ri}$ with the Higgs and lepton doublets,
\begin{align}
\mathcal{L} &= \mathcal{L}_{\rm SM} + \i\,\bar\nu_{Ri}\slashpartial \nu_{Ri} - \left[\frac{1}{2}M_R^{ij}\,\bar\nu^c_{Ri}\nu_{Rj} + Y_D^{\alpha i}\,\bar L_\alpha \widetilde H \nu_{Ri} + {\rm h.c.}\right]\,,
\label{eq:LnuSM}
\end{align}
where $(\alpha=e,\mu,\tau)$ and $(i,j=1,2,3)$ are flavor and family indices, respectively. After electroweak symmetry breaking (EWSB), the neutrino mass matrix $M_\nu$ can be written using the weak-basis states $\mathcal{N}=(\nu_L\,,\ \nu_R^c)^{\sf T}$,
\begin{align}
\mathcal{L}_{\rm mass}=-\frac{1}{2}\,\overline{\mathcal{N}^c}\,M_\nu\,\mathcal{N} + {\rm h.c.}\ ,\quad M_\nu=\begin{pmatrix}
0 & m_D^{\sf T}\\
m_D & M_R^\dagger
\end{pmatrix}\,,
\label{eq:Lmass}
\end{align}
where $m_D=Y_D^\dagger \tfrac{v}{\sqrt{2}}$ and $v\simeq246\GeV$. The neutrino mass matrix $M_\nu$ is diagonalized by a $6\times6$ unitary matrix $\mathcal{U}$,
\begin{align}
\mathcal{U}^{\sf T}\,M_\nu\,\mathcal{U} &= \begin{pmatrix}
m_l & 0\\
0 & M_h
\end{pmatrix}\,,\quad
m_l =\diag(m_{\nu_1},m_{\nu_2},m_{\nu_3})\,,\notag \\
M_h &=\diag(M_1,M_2,M_3)\,.
\end{align}
This matrix relates the weak-basis states with the Majorana mass-basis states $\nu\equiv\mathcal{N}_m+\mathcal{N}_m^c$, with $\mathcal{N}_m=(\nu_L'\,,\ \nu_R^c{}^\prime)^{\sf T}$, via $\mathcal{N}=\mathcal{U}\mathcal{N}_m$,
\begin{align}
\nu_L &= U_{aa}\,\nu_L' + U_{as}\,\nu_R^c{}'\,, \notag \\
\nu_R^c &= U_{sa}\,\nu_L' + U_{ss}\,\nu_R^c{}'\,,
\end{align}
where we have block-parameterized the matrix $\mathcal{U}$ as
\begin{align}
\mathcal{U} &= \begin{pmatrix}
U_{aa} & U_{as}\\
U_{sa} & U_{ss}
\end{pmatrix}\,.
\end{align}

Ref. \cite{Donini:2012tt} introduced a parametrization for $\mathcal{U}$  which is inspired by the usual Casas-Ibarra (C-I) 
approach \cite{Casas:2001sr} but does not assume an expansion in $m_lM_h^{-1}$,
\begin{align}
U_{aa} = U_\nu\,H\,, \quad\ \, U_{as} = \i\,U_\nu\,H\,\Theta\,,\quad U_{sa} = \i\,\overline H\,\Theta^\dagger\,,\quad U_{ss} = \overline H\,, \label{eq:Uas_def}
\end{align}
where we have introduced the matrix
\begin{align}
\Theta=m_l^{1/2}RM_h^{-1/2}\,,
\label{eq:Theta_def}
\end{align}
$U_\nu$ is a generic $3\times3$ unitary matrix, $R$ is a $3\times3$ complex orthogonal matrix, and $H,\overline H$ are $3\times3$ hermitian matrices determined by
\begin{align}
H &= \Big(1 + \Theta\Theta^\dagger\Big)^{-1/2}\,,\qquad \overline H = \Big(1 + \Theta^\dagger\Theta\Big)^{-1/2}\,.
\end{align}
Notice that one recovers the usual C-I parametrization at leading-order in the seesaw limit of $H,\overline H\simeq1$. In our parameter scan, we are in this regime, up to corrections of $\mathcal{O} (10^{-3})$.
The matrix $U_\nu$ approaches the Pontecorvo–Maki–Nakagawa–Sakata (PMNS) matrix with corrections of $\mathcal{O}(m_lM_h^{-1})$. 
Hence, at leading-order in the seesaw limit, $U_\nu$ can be determined from neutrino oscillation data in terms of three mixing angles $\theta_{12}$, $\theta_{23}$, and $\theta_{13}$, one Dirac CP-violating phase $\delta_{\rm CP}$, and two CP-violating Majorana phases $\alpha_1$ and $\alpha_2$. The active neutrino masses in $m_l$ are obtained by selecting the lightest neutrino mass $m_{\nu_{\min}}$ and the neutrino mass hierarchy, in conjunction with the mass-squared splittings $\Delta m_{21}^2$ and $\Delta m_{3\ell}^2$\,\footnote{In normal hierarchy (NH), $\ell=1$ and $\Delta m_{3\ell}^2\equiv m_3^2-m_1^2>0$, whereas in inverted hierarchy (IH), $\ell=2$ and $\Delta m_{3\ell}^2\equiv m_3^2-m_2^2<0$ \cite{Esteban:2016qun, Gambit:RHN}.}.

The rotation matrix $R$ is decomposed into a real and an imaginary exponential matrix, $R=R_{\rm re}\,R_{\rm im}$, with $R_{\rm re}=\exp(A)$ and $R_{\rm im}=\exp(\i\Phi)$. 
Both $A$ and $\Phi$ are $3\times3$ real, anti-symmetric matrices that we parameterize as $A=[a_{ij}]=[a_k\,\epsilon^{ijk}]$ and $\Phi=[\phi_{ij}]=[\phi_k\,\epsilon^{ijk}]$, where $\epsilon$ is the Levi-Civita tensor (with $\epsilon^{123}=+1$).

\section{Parameter scan and experimental constraints}
\label{sec:scan}
The independent parameters introduced in Sec. \ref{sec:model} define an 18-dimensional parameter space which we scan according to the ranges shown in Table \ref{tab:scan_params}. The parameters $\{p_i\}$ were randomly generated, following a uniform distribution in either $p_i$ or in $\log_{10}(p_i)$, depending on their respective ranges. The selected ranges for the active neutrino sector correspond to the $3\sigma$ confidence intervals provided by the NuFIT Collaboration \cite{Esteban:2016qun}. 
For the sterile neutrino masses we use $M_2\equiv M_1+\Delta M_{21}$ and $M_3\equiv M_1+\Delta M_{21}+\Delta M_{32}$ 
with $\Delta M_{21,32}>0$, ensuring the ordering $M_1< M_2 <M_3$.

\begin{table}[htb]
  \centering
  \begin{tabular}{lrrr}
    \toprule
    \textbf{Parameter} & \textbf{Range} & \textbf{Distribution}\\
    \midrule
    \multicolumn{3}{l}{\textit{Active neutrino sector}}\\
    $\theta_{12}$ [rad] & $[0.547684, 0.628144]$ & uniform \\
    $\theta_{23}$ [rad] & $[0.670206, 0.925025]$ & uniform\\
    $\theta_{13}$ [rad] & $[0.139452, 0.155509]$ & uniform\\ 
    $m_{\nu_{\min}}$ [eV] & $[10^{-5},0.05]$ & log-uniform \\
    $\Delta m^2_{21}$ $[10^{-5}\,\text{eV}^2]$ & $[6, 9]$ & uniform \\
    $\Delta m^2_{3\ell}$ $[10^{-3}\,\text{eV}^2]$ & $[\pm 2, \pm 3]$ & uniform \\
    $\delta_{\rm CP}$ [rad] & $[0,2\pi]$ & uniform\\
    $\alpha_1$, $\alpha_2$ [rad] & $[0,2\pi]$ & uniform\\ \\
    \multicolumn{3}{l}{\textit{Sterile neutrino sector}}\\
    $a_{1}$, $a_{2}$, $a_{3}$ & $[-10,10]$ & uniform\\
    $\phi_{1}$, $\phi_{2}$, $\phi_{3}$ & $[-10,10]$ & uniform\\
    $M_1$ [GeV] & $[10^{-6},10^4]$ & log-uniform\\
    $\Delta M_{21}$ [GeV] & $[10^{-7},10^4]$ & log-uniform\\
    $\Delta M_{32}$ [GeV] & $[10^{-7},10^4]$ & log-uniform\\
    \bottomrule
  \end{tabular}
  \caption{Ranges for the 18-dimensional parameter space adopted for the model scan, with `$+$' and $\ell=1$ (`$-$' and $\ell=2$) for the normal (inverted) hierarchy of the active neutrino masses. We have introduced $\Delta M_{ij}\equiv M_i-M_j$.}
  \label{tab:scan_params}
\end{table}

We use the open-source software package \textsc{GAMBIT} \cite{Gambit:intro} to explore the parameter space using the sampling algorithm \textsc{Diver} \cite{Gambit:Diver}, a differential evolution-based scanner. To maximize the parameter space exploration, our strategy included performing multiple scans, activating and deactivating the use of self-adaptive evolution, based on the $\lambda$jDE algorithm \cite{Gambit:Diver}, for sampling fine-tuned regions in high-dimensional parameter spaces \cite{Gambit:RHN}. In terms of convergence and best-fit stability, we have utilized different convergence thresholds ($10^{-6}$, $10^{-8}$, and $10^{-10}$) to eliminate the chances of premature convergence to false optima.

Regarding the experimental constraints,  
we build upon the $\nu$SM implementation provided in \textsc{GAMBIT}~\cite{Gambit:RHN}. 
Ref.~\cite{Gambit:RHN}  included active neutrino mixing constraints, indirect constraints, and direct searches for sterile neutrinos heavier than 
$\sim\!10\,{\rm MeV}$.
We update the constraints for which new experimental information has become available, and 
include additional input because we explore a broader range of sterile neutrino masses, down to $M_1 \sim 1 \, {\rm keV}$. 
The changes compared to Ref.~\cite{Gambit:RHN} are: 

\begin{enumerate}
\item We added the most stringent nuclear beta-decay direct limits on $|\mathcal U_{ei}|^2$ for sterile neutrinos with $M_i\lesssim 10^{-3}\,{\rm GeV}$ \cite{Bolton:2022pyf}.
\item We updated the experimental inputs and bounds for the elements of the Cabibbo-Kobayashi-Maskawa (CKM) matrix \cite{Cirigliano:2022yyo, FermilabLattice:2018zqv, FlavourLatticeAveragingGroupFLAG:2024oxs, PhysRevC.102.045501}.
\item We adopted the 95\% C.L. upper bound on the sum of the active neutrino masses from the DESI 2024 VI results \cite{DESI:2024mwx}, $\sum_i m_{\nu_i}^{\rm (NH)}\le 0.113\eV$ and $\sum_i m_{\nu_i}^{\rm (IH)}\le 0.145\eV$. 
\item We included the latest $R_K$ and $R_{K^\star}$ results from the LHCb Collaboration \cite{LHCb:2022vje}, which show consistency with the SM. We also updated the $W$ boson decay branching fractions with results from the CMS Collaboration \cite{CMS:2022mhs}, including the correlations between different leptonic channels.
\item 
Certain observables only receive direct contributions from sterile neutrinos as long as their mass is below a particular threshold. To take this into account, we implemented several thresholds, 
one of which was set to $\sim\!m_\mu$ (for muon and meson decays) and another $\sim\!m_W$ (for processes around the EW scale). 
\end{enumerate}

In summary, the physical configurations explored in our scan included both the normal (NH) and inverted (IH) hierarchies for the active neutrinos and considered sterile neutrino masses ranging from the keV scale up to the TeV scale 
(see Table \ref{tab:scan_params}). 
In principle, the parameter space for light sterile neutrinos is severely constrained by Big Bang Nucleosynthesis (BBN) \cite{Dolgov:2003sg}. 
Non-standard cosmological scenarios avoiding these constraints have been studied, for example, involving entropy production \cite{Fuller:2011qy} or adding axion-like particles \cite{Deppisch:2024izn} as an alternative decay channel. Motivated by these \textit{beyond-}$\nu$SM possibilities, we considered scenarios with and without the addition of BBN constraints.  
In practice, ignoring BBN constraints opens up the parameter space for lighter sterile neutrinos 
and we will show in Section~\ref{sec:results} the corresponding changes in the maximum likelihood regions.

\section{$\boldsymbol{\onbb}$ half-life beyond light active neutrinos}
\label{sec:onbb}

Since the main objective of our analysis is to study $\onbb$ decay in the general type-I seesaw model, in this section we briefly recall the state-of-the-art treatment of 
$\onbb$ in the presence of sterile neutrinos. We follow the 
effective field theory (EFT) approach of Ref.~\cite{Dekens:2024hlz} in which the contribution of an active or sterile neutrino of mass $m_i$ to the inverse half-life can be written as
\begin{align}
\left(T_{1/2}^{0\nu}\right)^{-1}=g_A^4 G_{01}\Bigg| \sum_{i=1}^{6}\frac{\mathcal U_{ei}^2 \,m_i}{m_e} A_\nu(m_i)\Bigg|^2\,,
\end{align}
where $g_A = 1.2754\pm 0.0013$ \cite{ParticleDataGroup:2024cfk} is the axial-vector charge of the nucleon, $G_{01,\,{\rm Xe}} = 1.5\times 10^{-14}$ yr$^{-1}$ and $G_{01,\,{\rm Ge}} = 0.22\times 10^{-14}$ yr$^{-1}$ are the phase-space factors \cite{Horoi:2017gmj} for $^{136}$Xe and $^{76}$Ge, respectively, and $A_{\nu}$ is the amplitude. The latter receives contributions from (sterile) neutrinos with different momentum scalings, depending on their masses,
\begin{align}
A_\nu (m_i) = \qquad\qquad\qquad\qquad\qquad\qquad\qquad\qquad\qquad\qquad\quad\notag\\
\label{eq:fullint}
\begin{cases}
A_\nu^{\rm (pot,<)}+A_\nu^{\text{(hard)}}(m_i)+A_\nu^{\text{(usoft)}}(m_i) \,,&  m_i <  100\MeV\,, \\
A_\nu^{\rm (pot)}+A_\nu^{\text{(hard)}}(m_i)\,,& 100\MeV \le m_i <  2\GeV\,,\\
A_\nu^{\text{(9)}}(m_i)\,,& {\rm }\,  2\GeV \le m_i \,.
\end{cases}
\end{align}

Here $A_\nu^{\rm (9)}$ captures contributions from sterile neutrinos that can be integrated out perturbatively (whenever $m_i\gtrsim 2\GeV$), in which case dimension-9 operators can effectively describe them.
$A_\nu^{\rm (hard)}$ captures the effects of hard neutrinos with momenta of $k_0\sim |\spacevec k|\sim 1$ GeV. In this momentum region, the effects are controlled by a short-distance interaction between four nucleons and its respective low-energy constant (LEC) $g_{\nu}^{N\!N}(m_i)$ \cite{Cirigliano:2024ccq}.
Instead, the superscripts $\rm (pot,<)$ and (pot) denote the contributions from neutrinos with potential momenta, $k_0\ll |\spacevec k|\sim m_\pi$.  
Here $A_\nu^{\rm (pot,<)}$ is equivalent to $A_\nu^{\rm (pot)}$ up to certain terms that would otherwise be double counted after combining with $A_\nu^{\rm (usoft)}$ in the $m_i<100$ MeV mass region.
The effects of these neutrinos are determined by Nuclear Matrix Elements (NMEs), similar to those induced by the long-range exchange of active neutrinos. 

Finally, $A_\nu^{\rm (usoft)}$ represents contributions from ultrasoft neutrinos with momenta of the order of the excited-state energies or the $Q$ value of the process, $k_0\sim|\spacevec k|\sim\Order({\rm MeV})$. These effects depend on transition matrix elements between the initial/final state and intermediate nuclear states, as well as their energies. The details of these contributions are discussed in Ref.~\cite{Dekens:2024hlz}.

The sensitivity of  $0\nu \beta \beta$ decay to  lepton-number-violating new physics is usually 
presented in terms of the parameter $m_{\beta \beta} = \sum_{i=1}^3 \mathcal U_{ei}^2\, m_{\nu_i}$, 
which controls the decay in generic (not necessarily Type-I)  high-scale seesaw scenarios
(\emph{i.e.,} $1/T_{1/2}^{0\nu}  \propto |m_{\beta \beta}|^2$). 
While in the generic Type-I seesaw model the $0\nu \beta \beta$ decay rate is not proportional to $|m_{\beta \beta}|^2$, in analogy with the high-scale seesaw, we define the effective Majorana mass \cite{Cirigliano:2018yza}:
\begin{align}
\mbb = \frac{m_e}{g_A^2 |V_{ud}|^2 \mathcal{M}_{\nu}^{(3)} G_{01}^{1/2}} \left( T_{1/2}^{0\nu} \right)^{-1/2}\,,
\label{eq:mbb_def}
\end{align}
which reduces to $m_{\rm \beta \beta}$ as $M_R$  grows significantly above the electroweak scale. 
In Eq.~\ref{eq:mbb_def}  $V_{ud}$ is the $u-d$ component of the CKM matrix and $\mathcal{M}_{\nu}^{(3)}=2.7\ (3.4)$ is the active, light neutrino NME \cite{Cirigliano:2018yza, Dekens:2024hlz} for the decay of $^{136}$Xe ($^{76}$Ge).

\section{Results and Discussion}
\label{sec:results}
In this section, we present the main results of the parameter scan covering the scenarios of normal (NH) and inverted (IH) hierarchies of active neutrino masses with and without the presence of BBN constraints using a frequentist statistical approach \cite{StatsBook}. 

The consistency of a given parameter point $\mathbf{X}$ with experimental data $D$ is quantified by the likelihood function, defined as $\mathcal{L}(\mathbf X)\equiv \mathbb P(D\,|\,\mathbf X)$ --the probability of observing the data $D$ given the model parameters $\mathbf X$. Although the likelihood function generally depends on multiple parameters, we will often present results focusing on a subset of them, denoted $\mathbf{x}$.
The dimensionality of the parameter space can be reduced by introducing the profile likelihood, defined as $\mathcal{L}(\mathbf{ x})\equiv\underset{\mathbf{ x^\prime}}{\max}\,\mathcal{L}(\mathbf { x},\mathbf { x^\prime})$, which involves maximizing the likelihood with respect to the 
complement of $\mathbf{x}$, called $\mathbf{x}^\prime$.
Our results are presented in terms of 1- or 2-dimensional profile likelihood ratios, 
$\Lambda^{(n)}=\mathcal{L}/\mathcal{L}_{\max}$ for $n=1,2$.
Finally, likelihood intervals-- and more generally likelihood regions-- can be interpreted as confidence intervals via Wilk's theorem \cite{Wilks:1938dza}, 
through the relation 
$-2\ln\Lambda^{(n)}\leq F^{-1}_{\chi^2_n}(1-\alpha)$, where $F^{-1}_{\chi^2_n}(x)$ is the inverse cumulative distribution function for a $\chi^2$ distribution with $n$-degrees of freedom and $(1-\alpha)$ is the confidence level. For example, the 2D profile likelihood values equivalent to 
$\sim68\%$ and  $\sim95\%$  C.L. are $0.32$ and $0.05$, respectively.

For $M_i\gtrsim\Order(1\GeV)$, the behavior of the active-sterile mixing elements $|\mathcal U_{\ell i}|$ (with $\ell=e,\mu,\tau$ and $i=1,2,3$) as a function of $M_i$ is qualitatively consistent with the results presented in Ref. \cite{Gambit:RHN}. We omit such plots for conciseness and refer the reader to the reference above. For light sterile neutrino mass, not previously included in Ref. \cite{Gambit:RHN}, the behavior of the mixing elements depends on the BBN constraints. If those constraints are imposed, the light sterile neutrino parameter region is excluded; if those constraints are ignored, this region opens up and even becomes preferred by the scan.

\begin{figure*}[htb]
\centering
\includegraphics[width=\linewidth,clip]{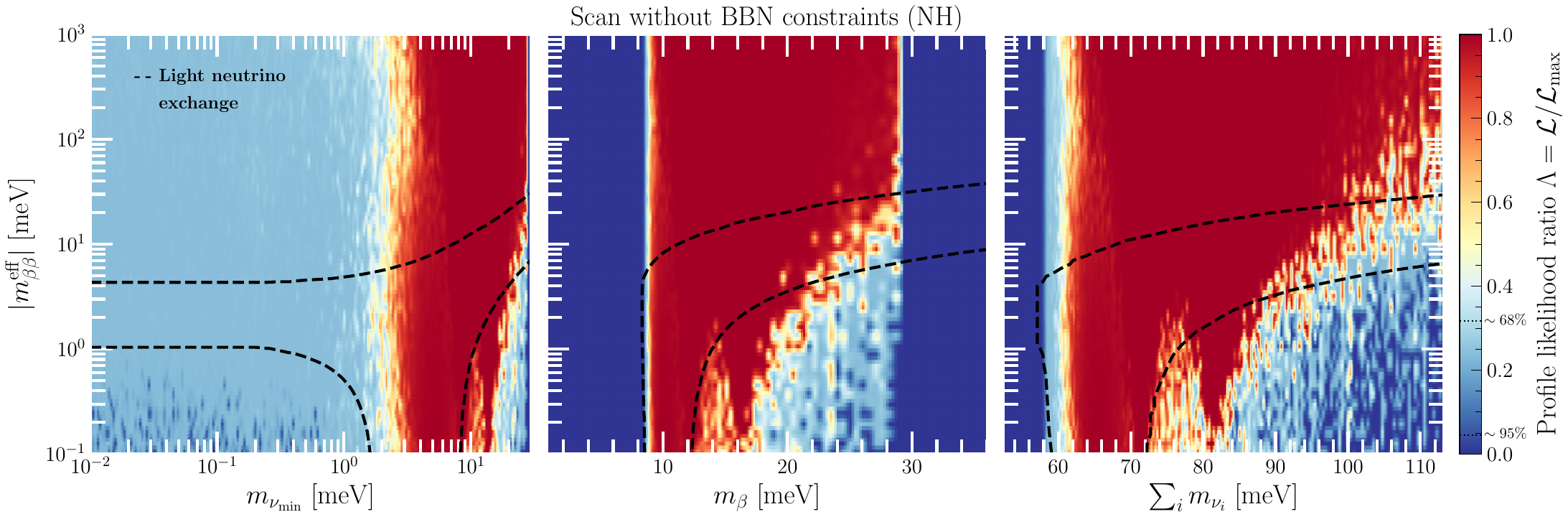}
\caption{2D profile likelihood of a scan setup without BBN constraints for $\mbb$ vs. $\mvmin$ (left), $\mb$ (center), and $\msum$ (right), respectively. The dashed black line corresponds to the bounds predicted by the light neutrino exchange mechanism ($\mbb \to |m_{\beta\beta}|$) in normal hierarchy (NH).}
\label{fig:mbb_BBNoff_NH}
\end{figure*}

\begin{figure*}[htb]
\centering
\includegraphics[width=\linewidth,clip]{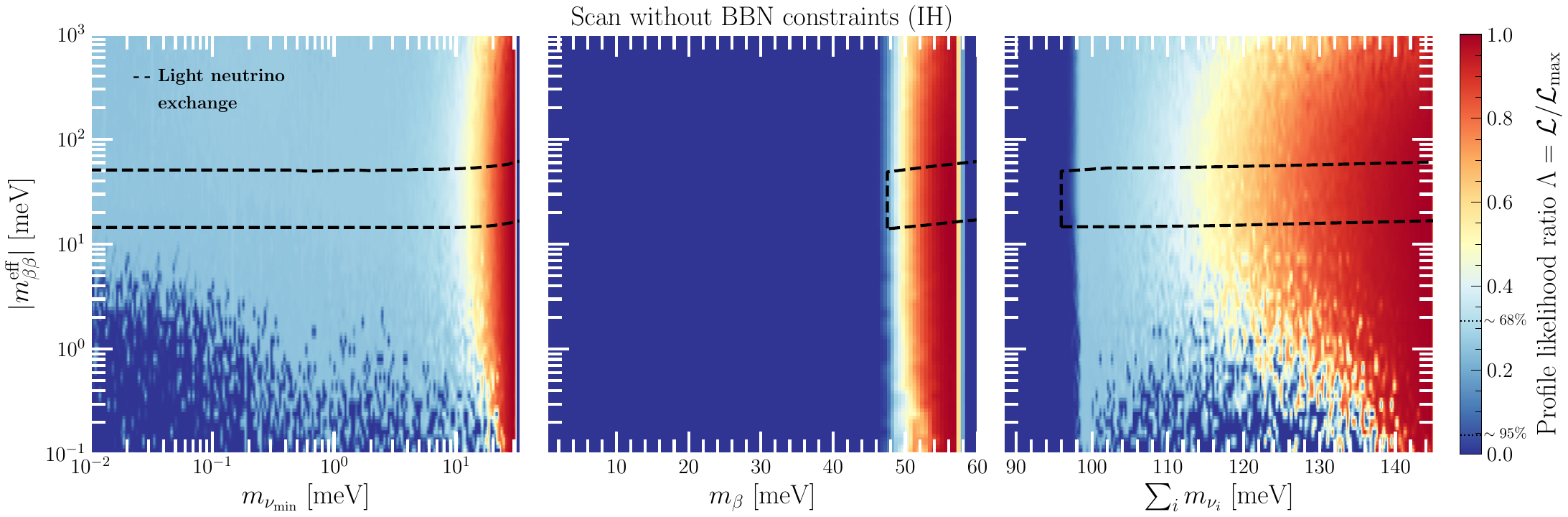}
\caption{Same as Fig. \ref{fig:mbb_BBNoff_NH}, but for inverse hierarchy (IH).}
\label{fig:mbb_BBNoff_IH}
\end{figure*}

\begin{figure*}[htb]
\centering
\includegraphics[width=\linewidth,clip]{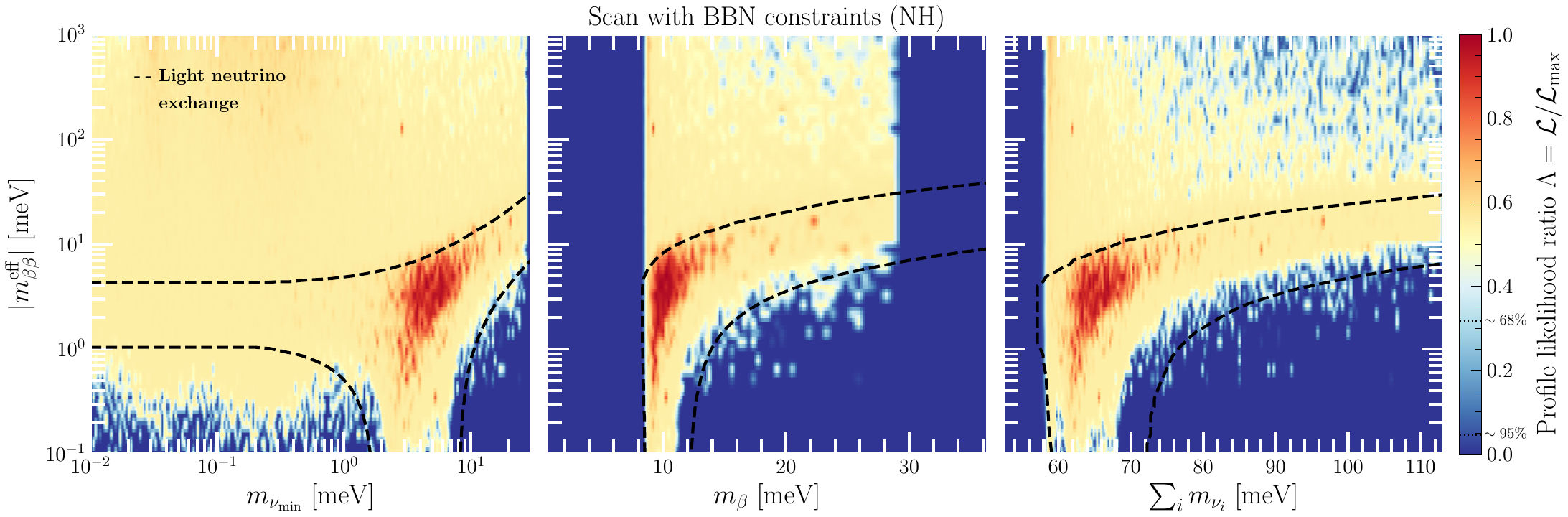}
\caption{2D profile likelihood of a scan setup imposing BBN constraints for $\mbb$ vs. $\mvmin$ (left), $\mb$ (center), and $\msum$ (right), respectively. The dashed black line corresponds to the bounds predicted by the light neutrino exchange mechanism in normal hierarchy (NH).}
\label{fig:mbb_BBNon_NH}
\end{figure*}

\begin{figure*}[htb]
\centering
\includegraphics[width=\linewidth,clip]{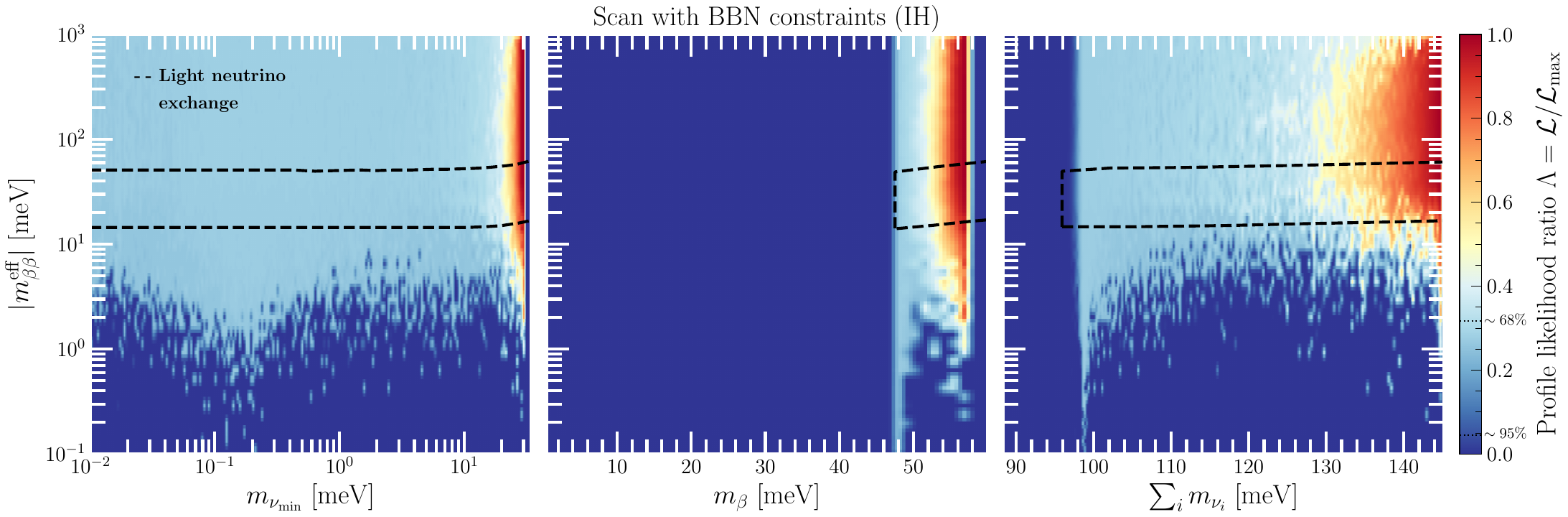}
\caption{Same as Fig. \ref{fig:mbb_BBNon_NH}, but for inverse hierarchy (IH).}
\label{fig:mbb_BBNon_IH}
\end{figure*}

We focus our analysis on the implications for $\onbb$. In Figs.~\ref{fig:mbb_BBNoff_NH}--\ref{fig:mbb_BBNon_IH}, we show the correlation of $\mbb$, defined 
in Eq.~\eqref{eq:mbb_def}, with the three light  neutrino mass combinations  
$\mvmin$, $\mb \equiv \sqrt{\sum_i|\mathcal U_{ei}|^2\,m_{\nu_i}^2}$ (constrained by Tritium Beta Decay experiments \cite{Project8:2022wqh, KATRIN:2022ayy}), and $\msum$, 
for NH and IH (with and without BBN constraints). 
The dashed lines in each panel 
show the allowed parameter space in the limit $|m_{\beta \beta}^{\rm eff}| \to |m_{\beta \beta}|$  (see, for instance, Ref. \cite{Esteban:2024eli}), \emph{i.e.,} the case in which sterile neutrinos are very heavy and decoupled, thus leaving the light Majorana neutrinos as the only contributors to $0\nu \beta \beta$ decay.


The salient features of Figs.~\ref{fig:mbb_BBNoff_NH}--\ref{fig:mbb_BBNon_IH}
can be understood through the following considerations. 
First, 
the kinematic lower limit $\mvmin \to 0$ implies a lower limit in both $\mb$ and $\msum$, depending on the mass hierarchy. Conversely, the cosmological upper bound on $\msum$ \cite{DESI:2024mwx} translates into an upper limit for $\mvmin\lesssim30\meV$ (for both NH and IH) and a corresponding upper limit for $\mb$ as well.  
The sharp drop in the likelihood at $\mvmin\sim 2\meV$ for NH and $\mvmin\sim20\meV$ for IH 
visible in Figs.~\ref{fig:mbb_BBNoff_NH} and \ref{fig:mbb_BBNoff_IH} (no BBN constraint) 
is determined by a combination of the electroweak precision observables (EWPO) considered in the analysis~\cite{Gambit:intro, Gambit:RHN}, where the sterile neutrinos are responsible for modifying the value of $G_F$ measured via muon decay by a factor of $f_{\rm ewpo}\equiv[1-(U_{as}U_{as}^\dagger)_{ee}-(U_{as}U_{as}^\dagger)_{\mu\mu}]$~\cite{Drewes:2015iva}. Given the 
current precision and tensions in the EWPO~\cite{ParticleDataGroup:2024cfk}, the  largest likelihood occurs in regions of parameter space with $f_{\rm ewpo}\neq1$, 
which requires non-zero active-sterile mixing, which in turn disfavors $\mvmin \to 0$ 
(see Eqs.~\eqref{eq:Uas_def} and \eqref{eq:Theta_def}). 
All in all, this leads to a preference for $\mvmin\gtrsim2\meV$ for NH and $\mvmin\gtrsim20\meV$ for IH.
Most of the features concerning bounds or favored values in 
$\mvmin$, $\mb$, and $\msum$ are also visible, albeit less prominently, 
in Figs.~\ref{fig:mbb_BBNon_NH} and \ref{fig:mbb_BBNon_IH}, with favored regions shrinking due to the additional BBN constraints. 

\begin{figure*}[htb]
\centering
\includegraphics[width=0.7\linewidth,clip]{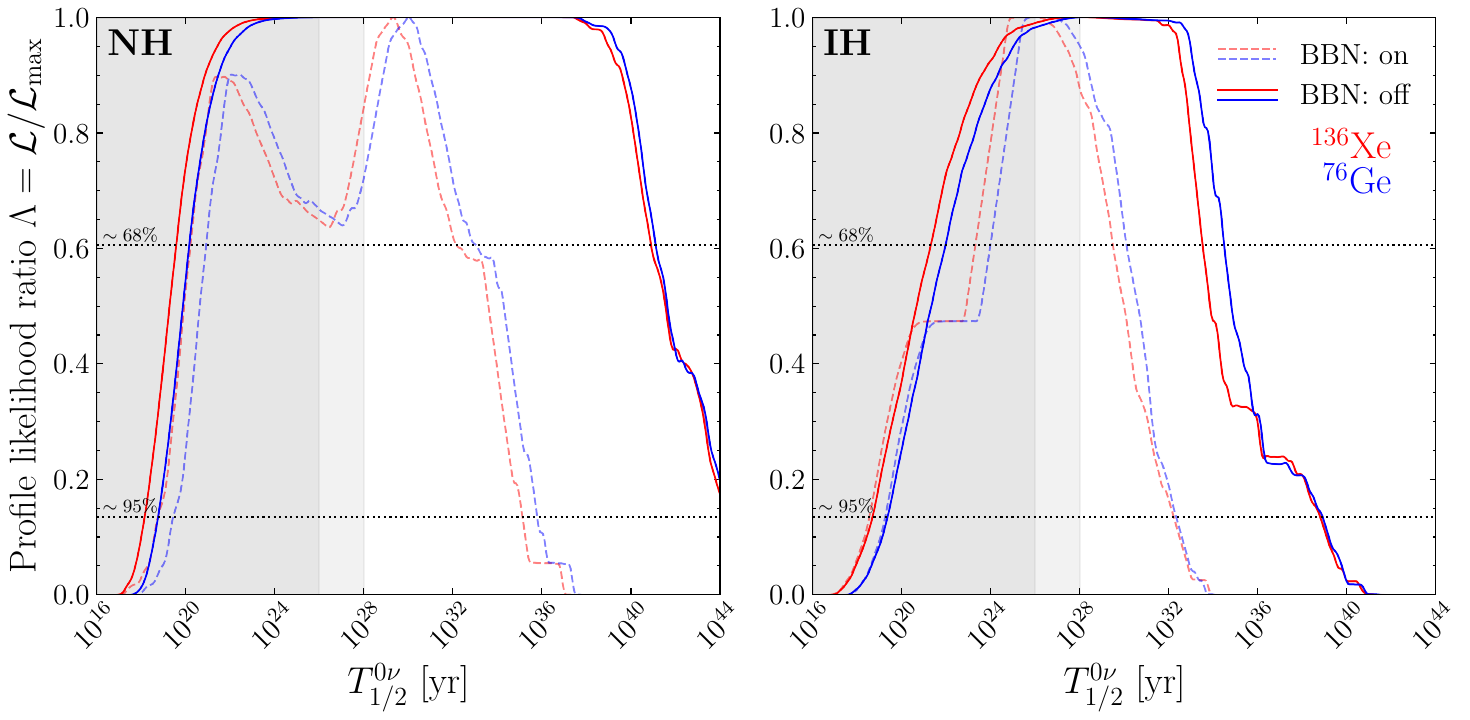}
\caption{1D profile likelihood for the half-life $T_{1/2}^{0\nu}$ of $^{76}{\rm Ge}$ (blue) and $^{136}{\rm Xe}$ (red) in NH and IH. The solid (dashed) lines represent scenarios without (with) BBN constraints. The gray-shaded regions indicate experimental limits: darker for the current experiments ($\sim10^{26}\,{\rm yr}$) \cite{GERDA:2020xhi, KamLAND-Zen:2022tow} and lighter future ton-scale experiments ($\sim10^{28}\,{\rm yr}$) \cite{Gomez-Cadenas:2019sfa, LEGEND:2017cdu, nEXO:2017nam, Han:2017fol, Armengaud:2019loe, Paton:2019kgy}.}
\label{fig:T120nubb}
\end{figure*}

Most importantly, in all four scenarios presented in Figs.~\ref{fig:mbb_BBNoff_NH}--\ref{fig:mbb_BBNon_IH}, the $68\%$ C.L. region in the profile likelihood extends well beyond that of the light neutrino exchange mechanism typically associated with a `high-scale seesaw' scenario (\emph{e.g.,} see Refs. \typeIrefs\ and references therein). 
{Qualitatively similar effects have also been observed in scenarios with $N=2$ sterile neutrinos \cite{deVries:2024rfh} (or more restricted parameter scans in the $N=3$ case \cite{Abada:2018oly}).}
This feature 
can also be seen from Fig.~\ref{fig:T120nubb}, where we present a 1-dimensional profile likelihood for the half-life $T_{1/2}^{0\nu}$ of two isotopes, $^{76}{\rm Ge}$ (blue) and $^{136}{\rm Xe}$ (red). The solid (dashed) lines represent the scenarios without (with) BBN constraints for both NH and IH. The gray-shaded regions depict experimental limits: darker for current experiments ($\sim10^{26}\,{\rm yr}$) \cite{GERDA:2020xhi, KamLAND-Zen:2022tow} and lighter for future ton-scale experiments ($\sim10^{28}\,{\rm yr}$) \cite{Gomez-Cadenas:2019sfa, LEGEND:2017cdu, nEXO:2017nam, Han:2017fol, Armengaud:2019loe, Paton:2019kgy}.   
The main features of Fig.~\ref{fig:T120nubb} can be understood as follows: 
(1) When the BBN constraint is not imposed, the profile likelihood saturates to $\sim 1$ for a broad range of  $T_{1/2}^{0\nu}$,  reflecting the wide range with high likelihood of $\mbb$ in Figs.~\ref{fig:mbb_BBNoff_NH} and \ref{fig:mbb_BBNoff_IH}; 
(2) When the BBN constraint is imposed,  Figs.~\ref{fig:mbb_BBNon_NH} and \ref{fig:mbb_BBNon_IH} 
illustrate a restricted region of high likelihood for $\mbb$, which implies a smaller range 
of $T_{1/2}^{0\nu}$ in which the profile likelihood saturates to $\sim 1$  
(in the case of NH  a secondary peak not visible in  Fig.~\ref{fig:mbb_BBNon_NH} appears at $\mbb\sim10^4\meV$, responsible for the peak in the profile likelihood in Fig.~\ref{fig:T120nubb} around $\Thalf\sim10^{20}\,{\rm yr}$); 
(3) For both NH and IH, the BBN constraint makes the profile likelihoods drop to 95\% C.L. 
for smaller $T_{1/2}^{0\nu}$ compared to the case without BBN constraint.  
This is because BBN disfavors the region with small sterile neutrino masses 
in which the $0\nu \beta \beta$ decay amplitude can be suppressed due to 
destructive interference of active and sterile contributions~\cite{Dekens:2024hlz}, 
thus resulting in a smaller upper bound on $T_{1/2}^{0\nu}$.

\begin{figure*}[htb]
\centering
\includegraphics[width=1.0\linewidth,clip]{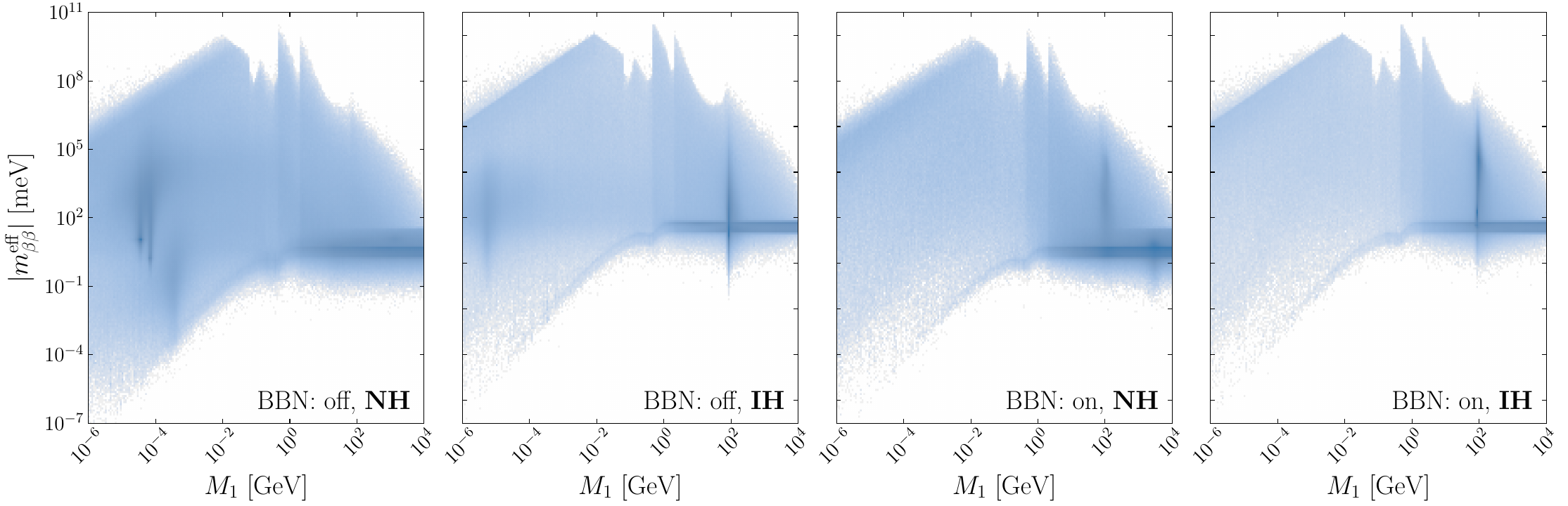}
\caption{2D projection of the scanned parameter space onto the $\mbb-M_1$ plane. The color intensity represents the point density.}
\label{fig:M10nubb}
\end{figure*}

Finally, to better understand the general features of Fig. \ref{fig:T120nubb}, we project 
the results of our scan 
onto the $\mbb-M_1$ plane, as shown in Fig.~\ref{fig:M10nubb}. 
{The explored parameter space is determined by the ranges established in Table \ref{tab:scan_params},
which lead to a finite range of $M_1$ and allowed values of $\mbb$. 
A future kink analysis of the beta spectrum of tritium decay~\cite{KATRIN:2024cdt} will limit the range of the former by imposing a lower limit on $M_1$. 
The $\mbb$ values, shown in Fig. \ref{fig:M10nubb}, 
correspond to the values of $\Thalf$ in Fig. \ref{fig:T120nubb} with nonzero profile likelihood.}
As expected, in the IH, the overall region is shifted to larger values of $\mbb$ compared to NH. The shape of the upper boundary corresponds to the sterile neutrino amplitude scaling $\sim1/(k_F^2+M_i^2)$, with $k_F\sim100\MeV$, 
explicitly showing the transition between the  long-range light neutrino exchange ($M_1\ll\Lambda_\chi$) and   
the short-range contributions described by a dimension-nine operator ($M_1\gg \Lambda_\chi$)~\cite{Dekens:2024hlz}.
The lower boundary for $M_1\gtrsim100\MeV$ is determined by the contribution to $\mbb$ from the active neutrinos. On the other hand, for $M_1<100\MeV$  cancellations between sterile and active neutrino contributions at the amplitude level can occur, so the minimal value of $\mbb$ decreases for smaller $M_1$.

\section{Conclusions}
\label{sec:conclusions}
We have presented a comprehensive study of the $3+3$ $\nu$SM, one of the simplest neutrino mass models invoking the type-I seesaw mechanism. 
This model has been extensively studied and motivates many experimental searches 
for additional neutral leptons~\cite{Abdullahi:2022jlv}. 
Our analysis focuses mainly on the implications for upcoming $\onbb$ decay searches. 
We explored the parameter space compatible with current experimental constraints from both low- and high-energy using the open-source software~\textsc{GAMBIT}.~\footnote{In the next decade, new experiments such as DUNE~\cite{DUNE:2025sjq} and SHiP~\cite{SHiP:2025ows}  will further probe the parameter space of the $\nu$SM considered here and it will 
be important to incorporate their impact on  $\onbb$ decay  searches.} 
We considered multiple physical scenarios, {including both neutrino-mass hierarchies as well as scanning with and without BBN constraints}, and  
derived profile likelihood for $T_{1/2}^{0\nu}$, the half-life of $0\nu \beta \beta$ decay for both the $^{76}$Ge and $^{136}$Xe isotopes.  
Our results in Figs.~\ref{fig:mbb_BBNoff_NH}--
\ref{fig:T120nubb} 
demonstrate that the 
95\%  C.L. regions for 
$\mbb$ or equivalently $T_{1/2}^{0\nu}$ 
extend well beyond the ones 
in the light neutrino exchange mechanism.  
Our findings imply that in this simple class of models for Majorana neutrino masses, current and next-generation $\onbb$ decay experiments have a broad discovery potential in both cases of normal and inverted ordering of the spectrum of light active neutrinos.

\section*{Acknowledgments}

We thank André de Gouvêa and Jordy de Vries for valuable discussions and Tomas Gonzalo for his technical support regarding \textsc{GAMBIT}. This work was supported by the INT’s U.S. Department of Energy grant No. DE-FG02-00ER41132.

\appendix

\bibliographystyle{apsrev4-1}
\balance
\biboptions{sort&compress}
\bibliography{seesaw}

\end{document}